# Superconductivity and antiferromagnetism in Fe(Te$_{1-x}$S$_x$)$_y$ system

M.H. Fang[1,2], B. Qian[1], H.M. Pham[3], J.H. Yang[2], T.J. Liu[1], E.K. Vehstedt[1], L. Spinu[3], and Z.Q. Mao[1]

[1]*Department of Physics, Tulane University, New Orleans, LA 70118, USA*

[2]*Department of Physics, Zhejiang University, Hangzhou 310027, China*

[3]*Advanced Materials Research Institute and Department of Physics, University of New Orleans, LA 70148, USA*

Abstract

We have synthesized polycrystalline samples and single crystals of Fe(Te$_{1-x}$S$_x$)$_y$, and characterized their properties. Our results show that the solid solution of S in this system is limited, < 30%. We observed superconductivity at ~ 9 K in both polycrystalline samples Fe(Te$_{1-x}$S$_x$)$_y$ with 0< $x$ ≤ 0.3 and 0.86 ≤ $y$ ≤ 1.0, and single crystals with the composition Fe(Te$_{0.9}$S$_{0.1}$)$_{0.91}$, consistent with the recent report of $T_c$ ~ 10 K superconductivity in the FeTe$_{1-x}$S$_x$ polycrystalline samples with $x$ = 0.1 and 0.2. Furthermore, our systematic studies show that the superconducting properties of this system sensitively depend on excess Fe at interstitial sites and that the excess Fe suppresses superconductivity. Another important observation from our studies is the coexistence of the superconducting phase and antiferromagnetism. Our analyses suggest that this phase coexistence may be associated with the random distribution of excess Fe and possibly occurs in the form of electronic inhomogeneity.

I. INTRODUCTION

The recent development in Fe-based superconductors is remarkable. After the report of superconductivity at 26 K in LaFeAsO$_{1-x}$F$_x$ [1], materials with much higher superconducting transition temperatures up to 43-56 K were immediately discovered in LaFeAsO$_{1-x}$F$_x$ under high pressure [2], and in Sm-, Ce-, Nd- or Gd- substituted isostructural systems [3-7]. Moreover, superconductivity has also been revealed in oxygen-free systems such as (Ba$_{1-x}$K$_x$)Fe$_2$As$_2$ [8-10] and Li$_{1-x}$FeAs [11-13]. One of the remarkable properties of these materials is that their undoped parent compounds show SDW-type antiferromagnetic (AFM) orders, which either follow, or are accompanied by structure transitions [14-16]. Charge carrier doping in these materials suppresses long-range AFM orders and induces superconductivity, suggesting that magnetic correlations play an essential role in mediating superconducting pairing.

Another class of binary Fe-based superconductor α-FeSe with $T_c$~ 10 K was recently discovered [17], and possibly unconventional [18], and its $T_c$ can be enhanced to 27 K by applying hydrostatic pressure [19]. Interestingly, this compound contains antifluorite planes which are isostructural to the FeAs layer found in the iron arsenide. Both band structure calculations and photoemission experiments show that the Fermi surface structure of FeSe is essentially similar to those of the FeAs-based compounds [20-22]. In order to determine if the superconductivity in FeSe is associated with magnetic correlations, we previously studied the evolution of superconductivity and the phase diagram of the ternary Fe(Se$_{1-x}$Te$_x$)$_{0.82}$ (0 ≤ $x$ ≤ 1.0) system [23]. We found an enhanced superconducting phase with $T_{c,max}$ ~ 14 K in the 0.3 < $x$ < 1.0 range. This superconducting phase is suppressed when the sample composition approaches the end member FeTe$_{0.82}$. Similar results were also obtained by other groups [24-26].

Neutron scattering studies performed by Bao *et al*. [27] using our samples revealed simultaneous structural and AFM transitions near 65 K for the non-superconducting FeTe$_{0.82}$. In contrast with commensurate AFM order with collinear magnetic structure in FeAs-based compounds, the AFM order seen in FeTe$_{0.82}$ is incommensurate and includes both linear and spiral components. With Se substitution for Te, this long-range AFM order evolve into short-range AFM correlations and superconductivity occurs, suggesting that the superconducting pairing in Fe(Se$_{1-x}$Te$_x$)$_{0.82}$ might also be associated with magnetic correlations as in FeAs-based compounds. Another interesting observation from neutron scattering studies is that the nonstoichiometric composition of Fe(Se$_{1-x}$Te$_x$)$_{0.82}$ originates from the presence of excess Fe at interstitial sites, rather than chalcogen deficiency and that excess Fe also participates in the AFM order. The magnetic wave vector and the AFM transition temperature depend on the amount of excess Fe. In FeTe$_{0.90}$, which has less excess Fe compared to FeTe$_{0.82}$, the AFM order becomes commensurate and the transition temperature $T_N$ is increased to 75 K.

In this paper, we focus on sulfur-substituted iron telluride, i.e. Fe(Te$_{1-x}$S$_x$)$_y$. Since S has a smaller ionic radius than Te, it should produce chemical pressure as the Se substitution for Te does. Our motivation is to examine if this substitution could suppress the commensurate AFM order in FeTe$_{0.90}$ and induce superconductivity, and study the role of the excess Fe. We have successfully prepared polycrystalline samples Fe(Te$_{1-x}$S$_x$)$_y$ with $0 < x \leq 0.3$ and $0.86 \leq y \leq 1.1$, and single crystals with the composition Fe(Te$_{0.9}$S$_{0.1}$)$_{0.91}$. We indeed observed superconductivity with $T_c \sim 9$ K in this system, consistent with the recent report of $T_c \sim 10$ K superconductivity in the polycrystalline samples FeTe$_{1-x}$S$_x$ with $x = 0.1$ and 0.2 [28]. Furthermore, our systematic studies suggest that the superconductivity in this system is inhomogeneous and might coexist with an AFM order. Importantly we find that the superconducting volume fraction depends on excess Fe at interstitial sites and that the excess Fe suppresses superconductivity.

II. EXPERIMENT

The polycrystalline samples with nominal compositions Fe(Te$_{1-x}$S$_x$)$_{0.90}$ ($x$=0.0, 0.05, 0.1, 0.2, 0.25, and 0.3) and Fe(Te$_{0.7}$S$_{0.3}$)$_y$ ($y$ = 0.86, 0.90, 1.00) were synthesized using a conventional solid state reaction method. The mixed powder of Fe (purity 99.95%), Te (purity99.99%) and S (99.9%) was first ground, pressed into pellets, then sealed in an evacuated quartz tube and sintered at 700°C for 24 hours. The sample was then reground, pressed into pellets, and sintered again at 700°C for 24 hours. Single crystals of Fe(Te$_{0.9}$S$_{0.1}$)$_{0.91}$ were grown using a flux method. The powder mixed with the ratio Fe:Te:S = 1:0.6:0.25 was first sealed in an evacuated quartz tube, then heated up to 920°C and kept at this temperature for 24 hours, and finally cooled to 400°C at a rate of 5 °C /minute.

Samples were characterized by X-Ray diffraction (XRD) using a Rigaku Dmax 2000 X-ray diffractometer. Sample compositions were analyzed using a scanning electron microscope (Hitachi S3400) equipped with an Energy Dispersive X-Ray Spectrometer (EDXS). Resistivity measurements were performed using a standard four-probe method in a Physical Property Measurement System (PPMS, Quantum Design). The magnetization was measured with a SQUID (Quantum Design).

III. RESULTS AND DISCUSSIONS

Figure 1 shows the X-ray diffraction patterns at room temperature for samples Fe(Te$_{1-x}$S$_x$)$_{0.90}$ with $x$=0.0, 0.05, 0.1, 0.2, 0.25, and 0.3. The samples with $x < 0.2$ exhibit high purity; all of the diffraction peaks in these samples can be indexed with tetragonal lattices with the space group *P4/nmm*. For samples with $x > 0.2$, however, in addition to the major tetragonal phase, an impurity phase, i.e, $\beta$-FeS$_2$ (denoted by asterisks in Fig. 1), occurs. The amount of this impurity phase increases with increasing S content, suggesting that the solid solution of S in the tetragonal lattice is limited. Figure 2 presents the lattice parameters *a* and *c* derived from fitting the diffraction peaks for each composition. Both *a* and *c* decrease drastically with increasing S content for $x < 0.1$, but tend to saturate for $x > 0.1$. This result, together with the fact of the presence of impurity phase for $x > 0.2$, suggests that the solid solution limit of S may be < 30%.

We have also analyzed the compositions of these samples using EDXS. The measured compositions only slightly deviate from the nominal compositions for $x < 0.2$. For example, for the nominal composition Fe(Te$_{0.9}$S$_{0.1}$)$_{0.90}$, the measured composition is Fe(Te$_{0.9}$S$_{0.1}$)$_{0.95}$. The difference between them is within the limits of error for EDX analysis, suggesting that the actual composition of these samples is close to the nominal composition. Nevertheless, for samples with $x > 0.2$, the difference between the measured compositions and nominal compositions is relatively larger. This supports that the solubility limit of S is < 30%.

Figure 3 shows the temperature dependence of resistivity, $\rho(T)$, for Fe(Te$_{1-x}$S$_x$)$_{0.90}$ with $x$=0.0, 0.05, 0.1, and 0.2. The resistivity data of the $x = 0.0$ sample, FeTe$_{0.90}$, was previously reported in our early work [23]; we include it here for comparison. The most striking feature of this sample is that it undergoes a remarkable transition near 75 K. At temperatures above 75 K its resistivity shows a slight upturn with decreasing temperature, while below 75 K it switches to a metallic temperature dependence. Neutron scattering studies revealed that this transition corresponds to simultaneous first-order structural and AFM transitions [27,29]. As mentioned above, this AFM order is commensurate. From the resistivity data of the samples with $x = 0.05$, 0.1, and 0.2, we find that the partial substitution of S for Te indeed has a remarkable effect on the AFM transition. The transition shifts down to about 40 K for the $x = 0.05$ sample. Interestingly, this transition is followed by a sharp drop at ~ 9 K. In the sample with $x = 0.1$, this drop becomes much more significant and the resistivity decreases to zero at about 4 K, suggesting the occurrence of superconductivity. While we do not observe any feature in resistivity associated with the AFM transition in this sample, our magnetization data shows that AFM ordering still exists even at the superconducting state (see below). The $x = 0.2$ sample exhibits a similar transition at ~9 K, but its resistivity does not drop to zero (300 K - 2 K). This is possibly due to the presence of the impurity phase.

In Fig. 4, we present the magnetic susceptibility data for $Fe(Te_{1-x}S_x)_{0.90}$; they were measured under a magnetic field of 30 Oe with zero-field-cooling (ZFC) histories. All of the samples with various S content exhibit magnetic anomalies near 125 K. This feature was also observed in the $Fe(Se_{1-x}Te_x)_{0.82}$ system with $x > 0.4$. Neutron scattering measurements reveal neither structural nor magnetic transitions associated with this anomaly [27]. Therefore the origin of this magnetic anomaly is yet to be determined. In the $x = 0$ sample, the 125 K anomaly is followed by a second transition at about 65 K, which corresponds to the aforementioned AFM transition. This transition shifts to lower temperatures and becomes broader for other samples containing S. This can be seen clearly in Fig. 5a and 5b, which plot the susceptibility and resistivity data together, for the $x = 0.05$ and 0.10 samples respectively. For the $x = 0.05$ sample, the second transition temperature, $T_N$, shifts down to 38 K where a remarkable peak in the resistivity is observed. In the $x = 0.1$ sample, $T_N$ appears to decrease to ~ 22 K, but we did not observe any anomaly in the resistivity near this temperature. When the temperature is decreased below 9 K where the resistivity drops to zero, no significant superconducting Meissner effect occurs; instead we only observed a slight decrease in the susceptibility. The $x = 0.05$ sample shows a similar response in the susceptibility when the resistivity drop occurs near 9 K. These results suggest that the superconducting phase has a small volume fraction in these samples, and that it appears to coexist with an antiferromagnetically ordered state. From the discussions we present below, the small superconducting volume fraction in $Fe(Te_{1-x}S_x)_{0.90}$ can be attributed to the presence of excess Fe, which seems to be unfavorable to the superconducting phase.

As noted above, magnetic properties of $FeTe_y$ are sensitive to the excess Fe [27]. The AFM order can be tuned from incommensurate- to commensurate-type when $y$ is increased from 0.82 to 0.90. Given that magnetic correlations play an essential role in mediating the superconducting pairing of Fe-based superconductors, different magnetic ground states in parent compounds should lead to different superconductivity in doped materials. Therefore, it is reasonable to expect different superconducting properties in the S-substituted $FeTe_y$ series with various $y$. In order to examine this anticipated property, we prepared polycrystalline samples of $Fe(Te_{0.7}S_{0.3})_y$ with $y = 0.86$, 0.90, and 1.00. When we prepared this group of samples, we had not yet realized that the solubility limit of S is less than 30%, so we did not chose the most appropriate S content. XRD analyses of these samples show that they contain a small amount of impurity phase $\beta$-$FeS_2$, similar to that seen in the $Fe(Te_{0.7}S_{0.3})_{0.90}$ sample (see Fig. 1). The temperature dependences of resistivity below 20 K for these samples is presented in Fig. 6a. They all show the same onset transition temperature at ~9 K as in the Fe(Te$_{1-x}$S$_x$)$_{0.90}$ samples with $x > 0$, but the transition breadth is remarkably different among them. In the $y = 0.86$ and 0.90 samples which should have more excess Fe, the superconducting transition is very broad and their resistivities do not drop to zero in the measured temperature range. While the $y = 1.00$ sample, which has no or a minimum amount of excess Fe because of its stoichiometric nominal composition, the superconducting transition is much sharper and the resistivity decreases to zero at ~ 2.50 K. This suggests that the samples with little or no excess Fe have larger superconducting volume fractions than those with the same Te/S ratio but more excess Fe. This was confirmed in our magnetization measurements, as shown in Fig. 6b which presents the normalized susceptibility up to 20 K for the $y = 0.86$, 0.90 and 1.00 samples. For the $y = 0.86$ and 0.90 samples which have more excess Fe, they hardly displays superconducting diamagnetism though their resistivities show a broad superconducting transition below 9 K. However, for the $y = 1.00$ sample, which has no or minimum excess Fe, it exhibits remarkable superconducting diamagnetism below 9 K. The difference in superconducting diamagnetism between these samples strongly supports the argument that excess Fe suppresses superconductivity. Thus the small superconducting volume fraction in our samples of Fe(Te$_{1-x}$S$_x$)$_{0.90}$ can be well understood. In addition, as in Fe(Te$_{1-x}$S$_x$)$_{0.90}$, we observed coexistence of superconductivity and antiferromagnetism in the $y = 0.86$, 0.90, and 1.00 samples.

Regarding the origin of the coexistence of superconductivity and antiferromagnetism seen in our samples, we think that there two possibilities. One possibility is that the phase coexistence exists in the momentum space. In fact, this has been observed in FeAs-based compounds in which superconductivity was found to coexist with the spin-density wave (SDW) [30-33]. In this scenario of phase coexistence, the same electronic state participates in two different orders at different temperatures. The second possibility is that there exists a microscopic electronic phase inhomogeneity (electronic phase separation). In this scenario, the S substitution for Te does not completely suppress the antiferromagnetism, and induce superconductivity, in some local areas; this leads to an inhomogeneous distribution of superconducting and AFM phases in real space. Our results obtained from single crystals seem to support that this is the more probable scenario.

We have made efforts to grow $Fe(Te_{1-x}S_x)_{0.90}$ single crystals using a flux method as stated above. We have obtained high-quality crystals from the batch grown using the starting material with the Fe:Te:S =1:0.6:0.25 ratio. Crystal dimensions are typically ~3 mm × 2 mm ×0.1 mm. The inset to Fig.7 shows an image of one typical crystal. The lamellar shape of the crystal, together with the fact that crystals can easily be cleaved, reflects the characteristics of the layered structure of this material. The EDXS analysis shows that the composition of the crystal is $Fe(Te_{0.9}S_{0.1})_{0.91}$, and that it is homogenous. The main panel of Fig. 7 shows the XRD pattern of the single crystal. The observed (00*l*) diffraction peaks confirm that the single crystal has the same tetragonal structure as the $Fe(Te_{0.9}S_{0.1})_{0.90}$ polycrystalline sample.

We have measured in-plane resistivity and magnetic susceptibility using single crystal samples. The data is presented in Fig. 8. The susceptibility displays a remarkably sharp peak at ~ 30 K, which corresponds to the AFM transition; this feature is much more significant than that found in the polycrystalline sample. Following this transition, a second transition occurs at ~ 9 K below which the susceptibility shows a steeper drop and the resistivty decreases sharply, consistent with the observation in the Fe(Te$_{0.9}$S$_{0.1}$)$_{0.9}$ polcrystalline sample, i.e. superconductivity occurs below 9 K. Nevertheless, as seen in Fig. 8, the resistivity of the single crystal sample does not drop to zero even when the temperature is decreased to 0.3 K, and the susceptibility does not drop to a negative value, as expected for a typical Meissner Effect for a superconductor. These facts suggest that even in the single crystal, the superconducting phase has a small volume fraction. Our XRD analyses using powdered single crystals show our crystals to be high purity; they do not include any secondary phase intergrowth. Therefore, the more reasonable interpretation for this superconducting phase with a small volume fraction is the previously mentioned electronic phase separation, i.e. the minor superconducting phase coexists with the major AFM phase. This is probably caused by inhomogeneous distribution of excess Fe. We have already demonstrated that excess Fe participates in the AFM order and suppresses superconductivity. As a result, in the areas with rich excess Fe, an AFM order dominates and no or weak superconductivity occurs, while in those areas with little or no excess Fe stronger superconductivity appears. Electronic phase separation has been recognized as an important phenomenon of strongly correlated electrons. It has been extensively discussed in underdoped high-Tc cuprates [34] and manganese oxides [35]. The electronic phase separation we discussed here may represent an unusual example.

Another interesting feature which should be noted in the susceptibility of the single crystal is that it exhibits an anomaly at ~ 130 K, the origin of which is still unclear as indicated above. In comparison with that of the polycrystalline sample, this anomaly is less obvious and shifted to a higher temperature. This difference may be due to a slight difference in excess Fe between the polycrystalline and single crystal samples, which could lead to different magnetic properties.

IV CONCLUSION

In summary, we have studied the properties of $Fe(Te_{1-x}S_x)_y$ compounds using polycrystalline samples and single crystals. We find that, unlike the $Te(Se_{1-x}Te_x)_{0.82}$ solid solution series, the solid solution of S in $Fe(Te_{1-x}S_x)_y$ is limited, < 30 %. We observed superconductivity at ~ 9 K in all S-substituted samples, consistent with the recent report of $T_c$~ 10 K superconductivity in $FeTe_{1-x}S_x$ polycrystalline samples. We have demonstrated that the volume fraction of this superconducting phase is sensitive to the Fe excess and that the excess Fe suppresses superconductivity. In addition, our results reveal that the superconducting phase coexists with antiferromagnetism. This phase coexistence may be attributed to random distribution of the excess Fe.

**Acknowledgements**：The work at Tulane is supported by the NSF under grant DMR-0645305, the DOE under DE-FG02-07ER46358, the DOD ARO under W911NF-08-C-0131, and the Research Corporation; at Zhejiang University by National Basic Research Program of China (No.2006CB01003, 2009CB929104) and the PCSIRT of the Ministry of Education of China (IRT0754); at UNO is by DARPA through Grant No. HR0011-07-1-0031.

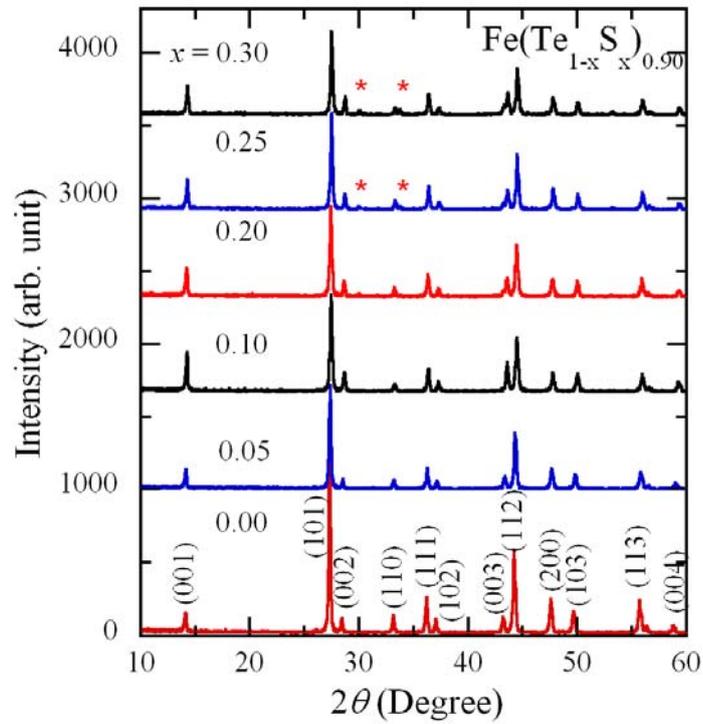

Figure 1: X-ray diffraction patterns of the polycrystalline samples with nominal composition Fe(Te$_{1-x}$S$_x$)$_{0.90}$. Peaks marked by "*": β-FeS$_2$ phase.

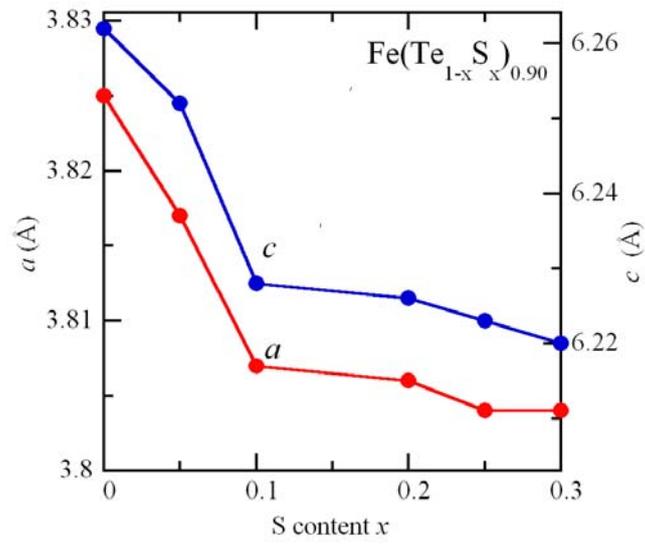

Figure 2: Lattice parameters, *a* and *c* as a function of S content *x* in the polycrystalline sample Fe(Te$_{1-x}$S$_x$)$_{0.90}$ series.

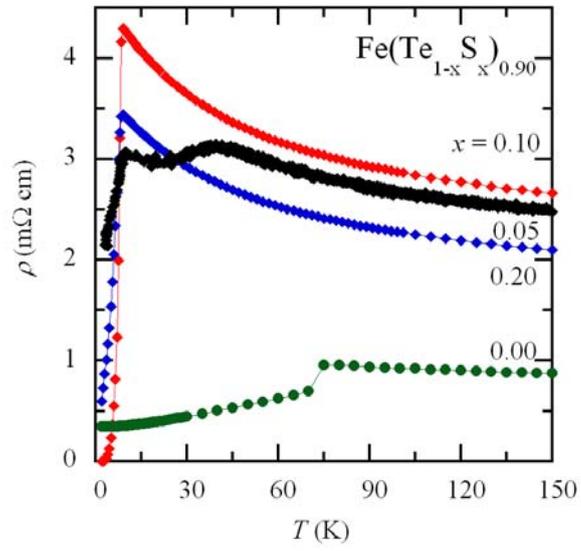

Figure 3: Resistivity as a function of temperature for the polycrystalline samples Fe(Te$_{1-x}$S$_x$)$_{0.90}$ ($x$=0.00, 0.05, 0.1 and 0.2).

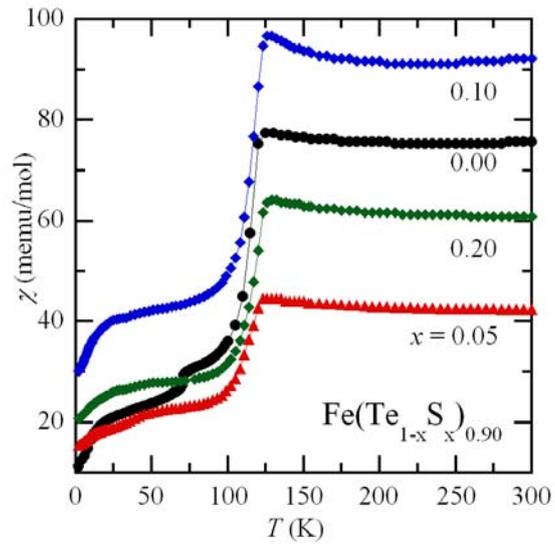

Figure 4: Temperature dependence of susceptibility measured under the field of 30 Oe with ZFC history for the Fe(Te$_{1-x}$S$_x$)$_{0.90}$ ($x$=0.00, 0.05, 0.10 and 0.20) polycrystalline samples.

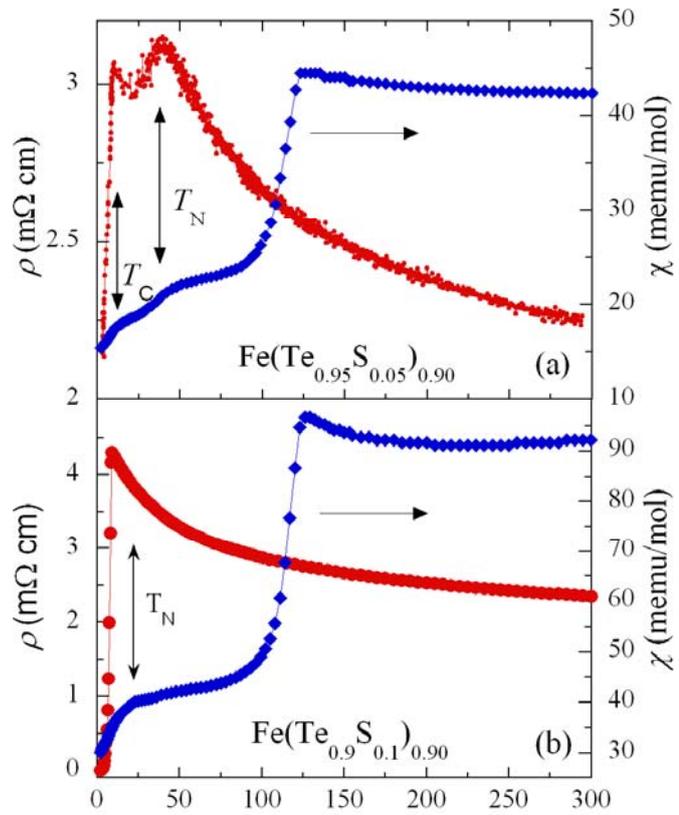

Figure 5: Resistivity and susceptibility as a function of temperature for two polycrystalline samples; (a) Fe(Te$_{0.95}$S$_{0.05}$)$_{0.90}$, (b) Fe(Te$_{0.9}$S$_{0.1}$)$_{0.90}$.

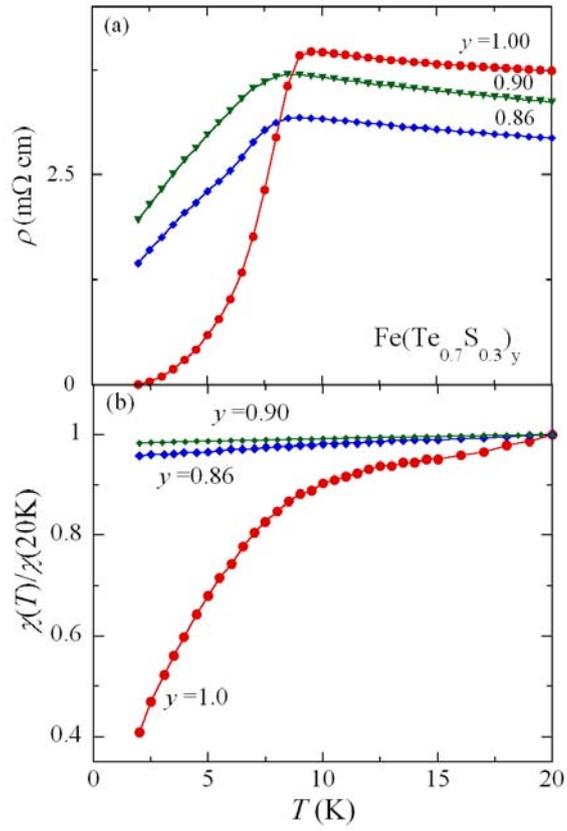

Figure 6: (a) Resistivity as a function of temperature for $T < 20$ K for Fe(Te$_{0.7}$S$_{0.3}$)$_y$; (b) Normalized susceptibility as a function of temperature for $T < 20$ K for Fe(Te$_{0.7}$S$_{0.3}$)$_y$.

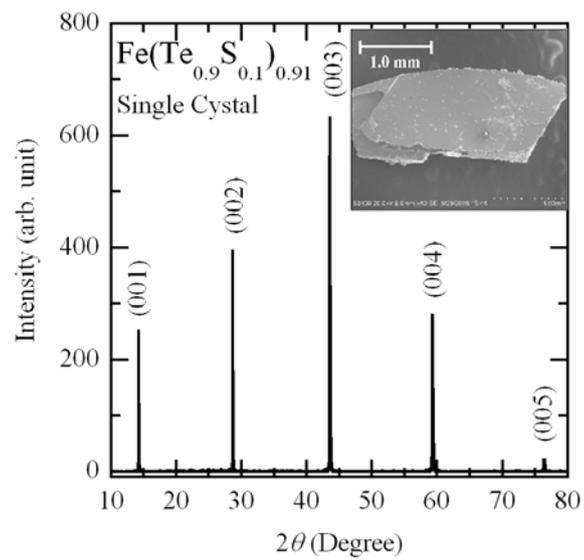

Figure 7: X-ray diffraction pattern of a typical Fe(Te$_{0.9}$S$_{0.1}$)$_{0.91}$ single crystal. The inset: The image of the Fe(Te$_{0.9}$S$_{0.1}$)$_{0.91}$ single crystal.

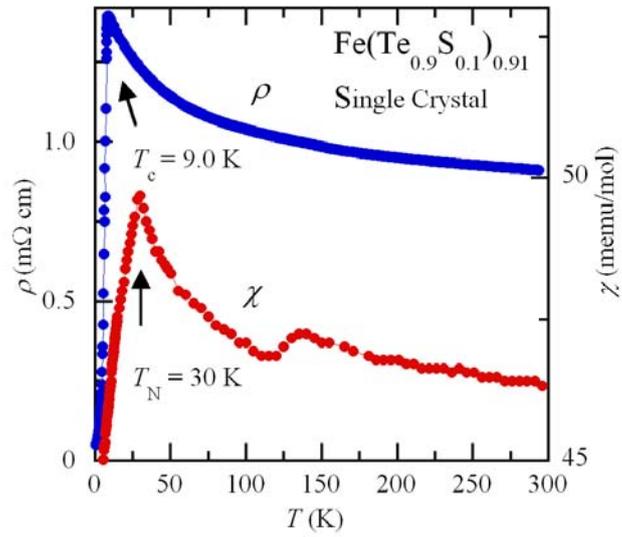

Figure 8: Resistivity and susceptibility as a function of temperature for Fe(Te$_{0.9}$S$_{0.1}$)$_{0.91}$ single crystals.